# New ways of scientific publishing and accessing human knowledge inspired by transdisciplinary approaches

I.C. Gebeshuber[a,b*] and B.Y. Majlis[a]

[a] Institute of Microengineering and Nanoelectronics, Universiti Kebangsaan Malaysia, 43600 UKM Bangi, Selangor, Malaysia

[b] Institute of Applied Physics, Vienna University of Technology, Wiedner Hauptstraße 8–10/134, 1040 Wien, Austria

* Corresponding author. Tel.: +60 12 392 9233; FAX: +60 3 8925 0439.
E–mail address: gebeshuber@iap.tuwien.ac.at (I.C. Gebeshuber)

## Abstract

Inspired by interdisciplinary work touching biology and microtribology, the authors propose a new, dynamic way of publishing research results, the establishment of a tree of knowledge and the localisation of scientific articles on this tree. 'Technomimetics' is proposed as a new method of knowledge management in science and technology: it shall help find and organise information in an era of over–information. Such ways of presenting and managing research results would be accessible by people with different kinds of backgrounds and levels of education, and allow for full use of the ever–increasing number of scientific and technical publications. This approach would dramatically change and revolutionize the way we are doing science, and contribute to overcoming the three gaps between the world of ideas, inventors, innovators and investors as introduced by Gebeshuber, Gruber and Drack in 2009 for accelerated scientific and technological breakthroughs to improve the human condition.

Inspiration for the development of above methods was the fact that – generally - tribologists and biologists do not see many overlaps of their professions. However, both deal with materials, structures and processes. Tribology is omnipresent in biology and many biological systems have impressive tribological properties. Tribologists can therefore get valuable input and inspiration from living systems. The aim of biomimetics is knowledge transfer from biology to technology and successful biomimetics in tribology needs collaboration between biologists and tribologists. Literature search shows that the number of papers regarding biotribology is steadily increasing. However, at the moment, most scientific papers of the other respective field are hard to access and hard to understand, in terms of concepts and specific wording, hindering successful collaboration and resulting in long times that are needed to speak and understand the other's language. For example, there is a plenitude of biology papers that deal with friction, adhesion, wear and lubrication that were written solely for a biology readership and that have high potential to serve as inspiration for tribology if they were available in a language or in an environment accessible for tribologists. The three needs that can be identified regarding successful biomimetics for microtribologists (i.e., joint language, joint way of publishing results, and joint seminars, workshops and conferences) are developed further into a general concept concerning the future of scientific publications and ordering as well as accessing the knowledge of our time.

Keywords: Biotribology; Biomimetics; Knowledge Transfer; Microtribology, Tree of Knowledge; Technomimetics; Publications; Organizing Knowledge

## 1. Introduction

Tribologists increasingly get interested in biology [1–8]. Just some years ago, the word “biotribology” was still mainly used to refer to investigations concerning hip or knee implants [9], i.e., to the tribology of prosthetic joints. Current biotribological research, however, increasingly deals with systems as diverse as the switchable attachment of blood cells to the endothelium [10,11] and the sticking of gecko feet to walls [12] (research that has inspired novel functional adhesives [13–15]), insect attachment pads and the respective structured plant surfaces [16,17] (research that might inspire novel mechanical attachment devices [18], imagine noiseless Velcro!), diatom hinges and interlocking devices [19] (research that might inspire emerging three–dimensional microelectromechanical systems [20–22]) and plant surfaces concerning their self–cleaning and anti–wetting properties [23,24] (for stain resistant paints and car, shoe or windshield coatings [25,26])

Biological systems are hierarchical, and in many cases, the micro– and nanoscale are crucial. Therefore, biomimetics seems to be especially promising in the field of micro– and nanotribology. The work by Scherge and Gorb who introduced biological micro– and nanotribology to engineering [1,27,28] has had a huge impact on the field and has been cited in nearly 200 scholarly works (Google Scholar search, performed in January 2010).

Literature search in the ISI Web of Knowledge, Science Citation Index Expanded (SCI–EXPANDED)—2001–present, reveals an increasing amount of scientific papers that link biology to tribology (Figure 1). And there is still a large, rather unexplored body of knowledge in biology that is relevant to tribology and that has not yet been linked extensively to technology (as example, see Figure 2 for biology articles with relation to wear and adhesives/adhesion).

In this paper, a new method is proposed to establish a tree of knowledge and a new way of scientific publishing that is accessible by people with different kinds of backgrounds and levels of education, to make full use of the ever–increasing number of scientific and technical publications. The example is given for linking biology and tribology papers, however, the general concept can be applied to all fields of science and technology, where trans– and interdisciplinary approaches are envisaged and where a tree of knowledge shall be established.

Already in 1751 Denis Diderot and Jean le Rond d'Alembert introduced a taxonomy of human knowledge (for English translation, see Figure 3) that served as table of contents for their systematic dictionary of the sciences [29].

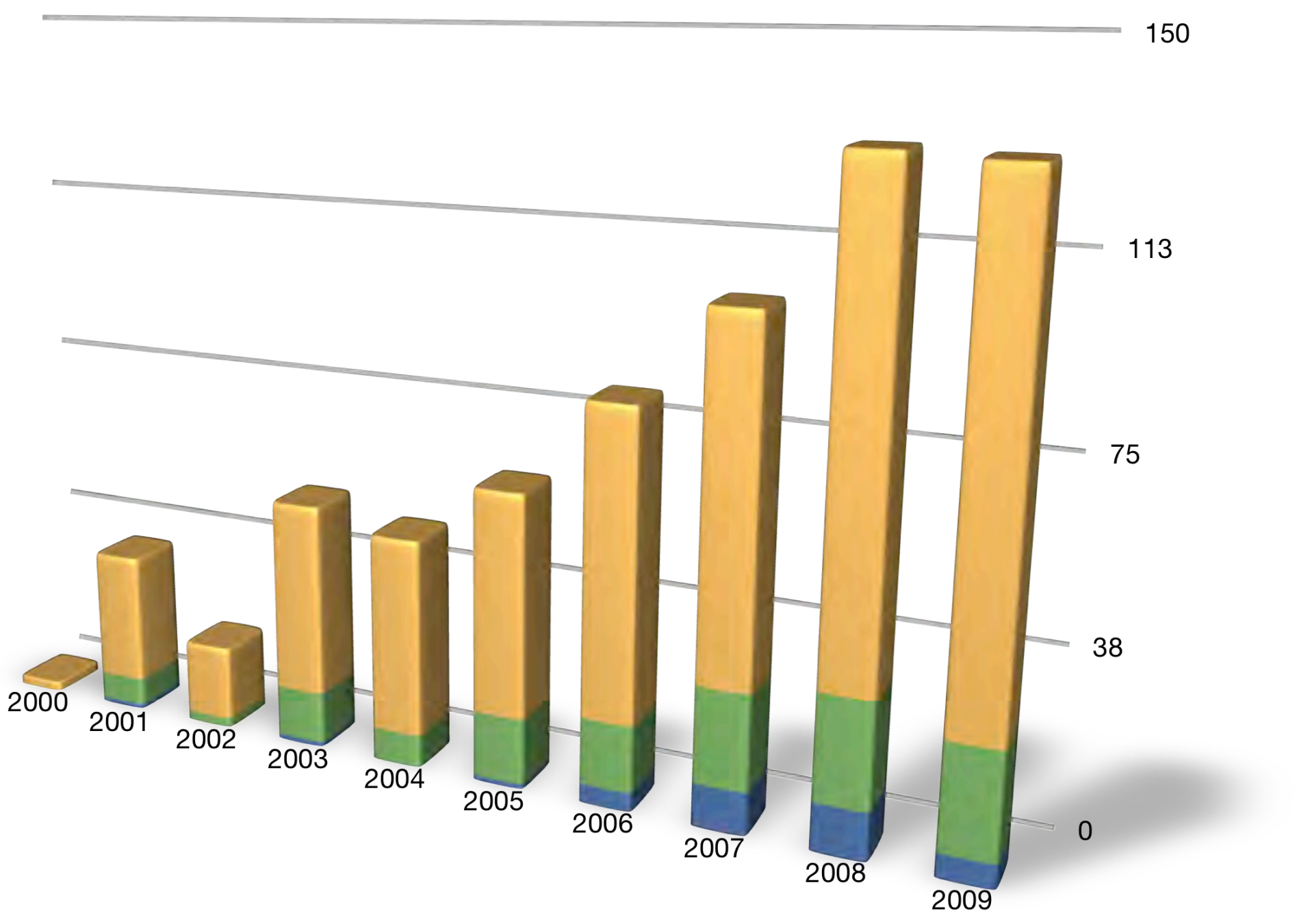

Figure 1. The number of scientific publications in the years 2000–2009 with explicit relation between biology and tribology (blue: 'biomim* and tribolog*' in topic, green: 'biolog* and tribolog*' in topic, yellow: 'bio* and tribolog*' in topic). Source: ISI Web of Knowledge, Science Citation Index Expanded (SCI–EXPANDED)—2001–present, Thomson Reuters. http://www.isiknowledge.com [accessed January 5, 2010]

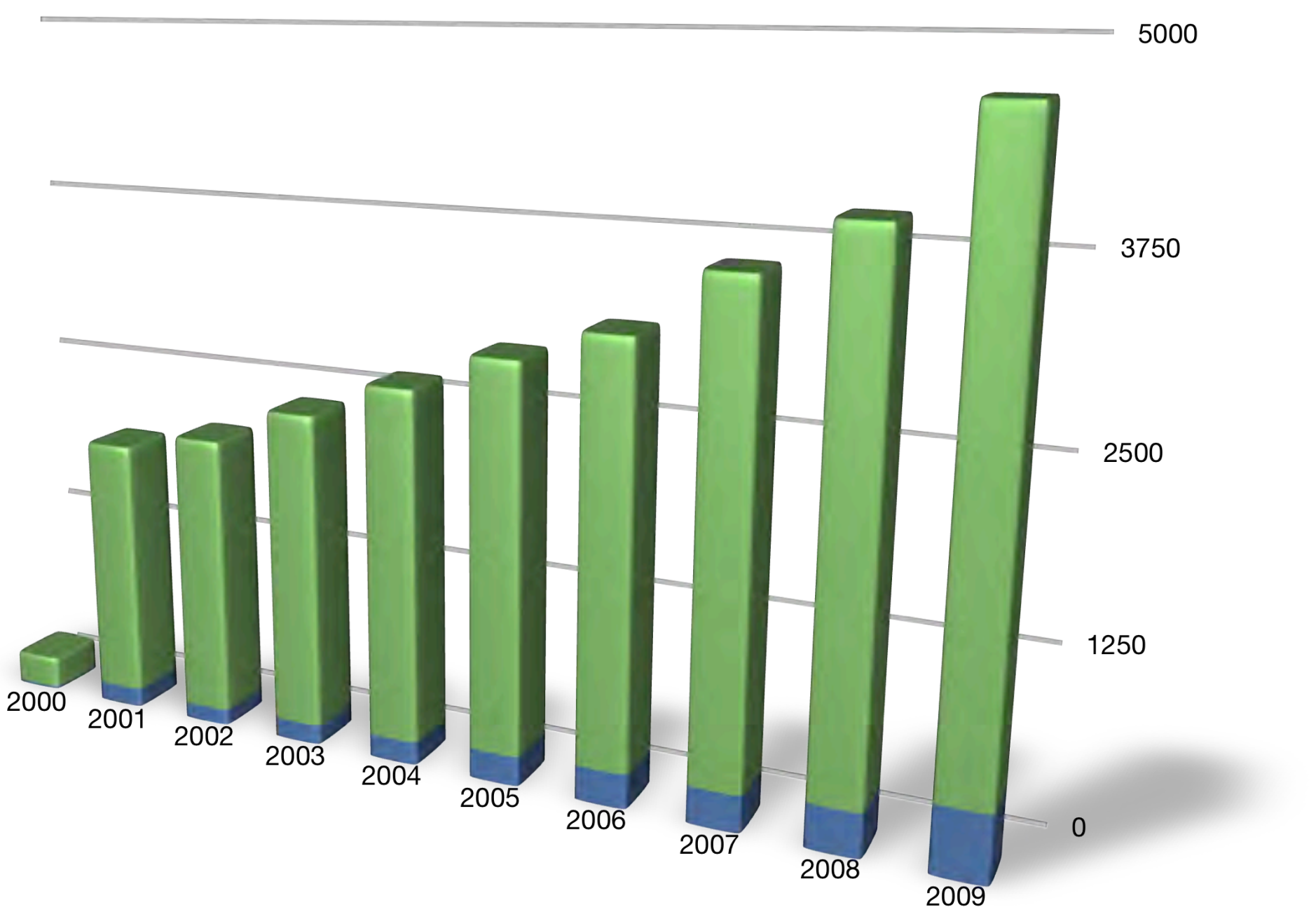

Figure 2. The number of scientific publications in the years 2000–2009 dealing with either wear or adhesives/adhesion in biology (blue: ‘bio* and wear’ in topic, green: ‘bio* and adhes*’ in topic) comprises a huge yet unexplored amount of inspiration for technology. Source: ISI Web of Knowledge, Science Citation Index Expanded (SCI–EXPANDED)—2001–present, Thomson Reuters. http://www.isiknowledge.com [accessed January 5, 2010]

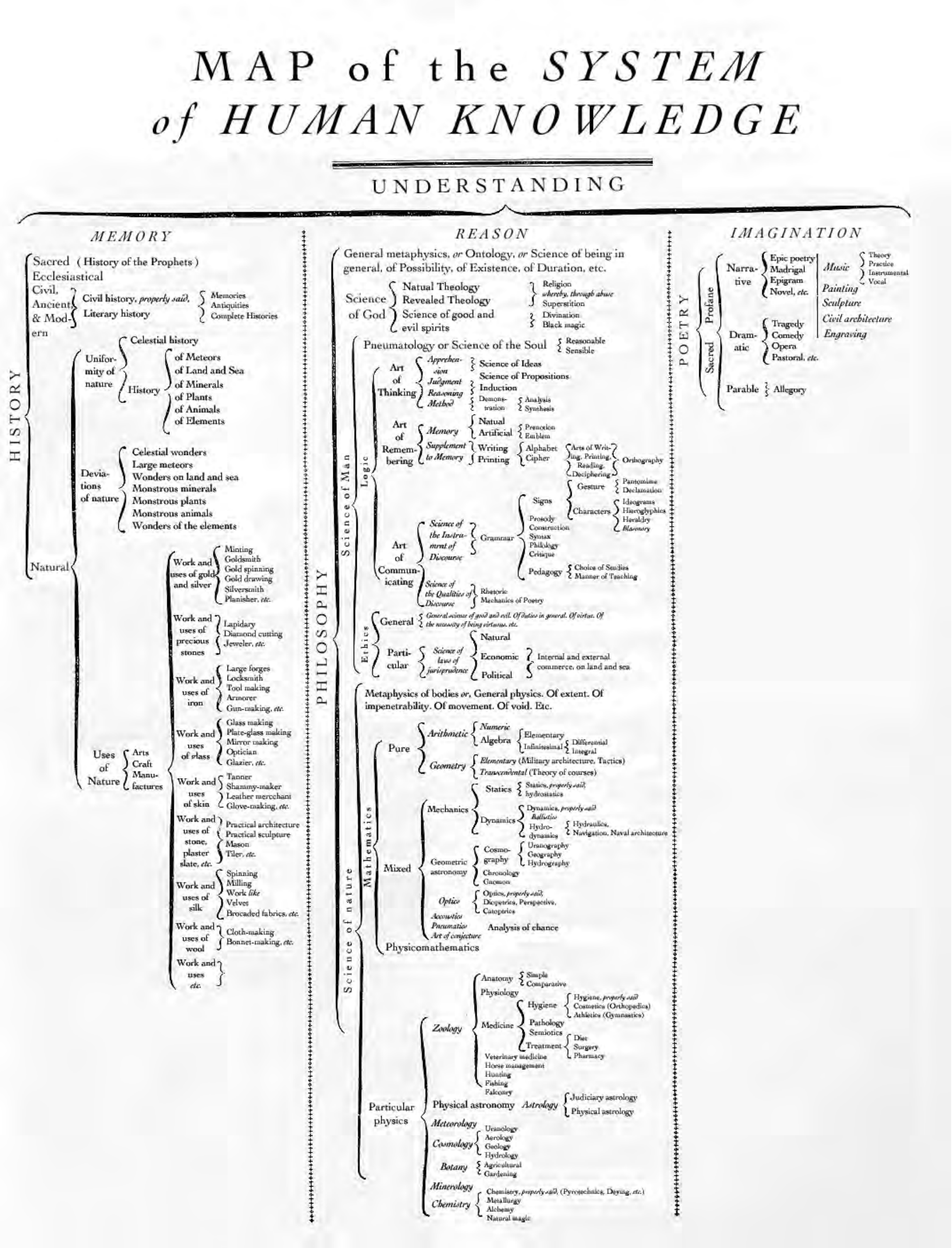

Figure 3. English translation of the “Tree of Knowledge” by Diderot and d’Alembert. © Benjamin Heller, University of Oxford, The Encyclopedia of Diderot and d'Alembert Collaborative Translation Project, http://quod.lib.umich.edu/d/did/tree.html. Image reproduced with permission.

The Biomimicry Innovation Method (BIM) [30] is a successful method in biomimetics that helps make use of biological literature and animated nature *per se*.

The BIM steps are "identify function", "biologise the question", "find nature's best practices" and "generate process/product ideas".
**Identify Function:** The challenges posed by engineers/natural scientists/architects and/or designers are distilled to their functional essence.
**Biologize the Question:** In the next step, these functions are translated into biological questions such as "How does nature manage lubrication?" or "How does nature manage joining rigid parts in relative motion?" The basic question is "What would nature do here?"
**Find Nature's Best Practices:** Scientific databases as well as e.g. scientific expeditions to the rainforest with its high species variety are used to obtain a compendium of how plants, animals and ecosystems solve the specific challenge.
**Generate Process/Product Ideas:** From the best practices, ideas for products and processes are generated.

BIM proves highly useful in habitats with high species variety and therefore high innovation potential (e.g., in the tropical rainforests), providing a multitude of natural models to learn from and emulate. Scientific expeditions to the Amazon as well as to Costa Rican and Malaysian pristine and secondary rainforests by one of the authors (ICG) kick–started biomimetic research in areas such as low–noise novel aircraft design (collaboration with Boeing engineers) and bioinspired technological development of novel reflectors and structural colours.
Gebeshuber and coworkers presented in 2009 an application of the Biomimicry Innovation Method concerning wear, shear, tension, buckling, fatigue, fracture (rupture), deformation and permanent or temporal adhesion, yielding a variety of best practices that comprise biological materials and processes in organisms as diverse as kelp, banana leafs, rattan, diatoms and giraffes [31]. Related to microtribology, these authors introduced a novel way to describe the complexity of biological and engineering approaches depending on the number of different base materials: Either many materials are used (*material* dominates) or few materials (*form* dominates) or just one material (*structure* dominates). They state that the complexity of the approach (in biology as well as in engineering) increases with decreasing number of base materials. In MEMS and NEMS technology, as in biology, a limited number of base materials is used, providing a wide range of functional and structural properties. The authors propose that because of material constraints in both fields, biomimetics is especially promising in MEMS development [21]. In a subsequent paper, they applied BIM regarding the innovation potential of biomimetics for novel 3D micro– and nanoelectromechanical systems (MEMS and NEMS) and present 20 different functions regarding structure dominated components along with biologised questions, natures best practices and generated product and process ideas, with 40 references to biological and technical literature [22].

Based on the knowledge gain that resulted from the preparation of these publications, the current publication develops ideas for a system that allows dealing with the ever–increasing amount of scientific publications in specialist areas and how to make them accessible for interdisciplinary approaches.

A major current problem in science is the huge amount of papers published. No single researcher can currently keep up with the reading even in his or her specialist field. Jack Sandweiss, the Editor of Physical Review Letters, dealt in an editorial address early in 2009 with the future of scientific publishing [32]. Sandweiss states "*For example, it is currently impossible for anyone to read all of Physical Review Letters or even to casually browse each issue.*" Note that Sandweiss here refers to one single journal only! He subsequently stresses the possibility of virtual journals, which collect links to articles in a particular field from many different journals and introduces an interesting new concept: "*a computer program that will be able to 'interview' scientists to find out what articles he or she would have selected if he or she had indeed been able to read all of the scientific literature. The interview would identify the number of articles each physicist would select and his or her personal priorities with respect to subfields. Such an artificial intelligence (AI) program would then read all the literature and select the relevant articles to provide to the subscriber*" [32]. In fact, such AI programs are already available, not in science though: in major online book, movie or music stores, expert systems that are based on ratings given by other users recommend the buyer further products. Depending on the history of the buyer known to the system the recommendations can be of high relevance and interest.

## 2. Biomimetics for Microtribology – Challenges and Opportunities

General equations and causal relations dominate our current engineering world. Biology papers are frequently inaccessible for engineers, since they are too descriptive and contain concepts and approaches such as taxonomy with its Latin names that are too far from any concept in engineering. A new type of presentation of knowledge is needed, perhaps even a new type of science, so that researchers from various fields can profit from each other findings. Linking different disciplines is not easy, however, transdisciplinarity has been widely suggested as an effective way of problem solving [see e.g., 33–35].

There is an abundance of biological literature available. However, only few of these works concentrate on the functions of biological materials, processes, organisms and systems [1,36–42].

The US American Biomimicry Guild has noticed the need for a tree of knowledge in biological literature that is sorted by function rather than by species or habitat. The Biomimicry Guild is currently undergoing a major endeavour and presents ‘strategies of nature’ together with scientific references and envisaged and already existing bioinspired applications in industry on its web–page http://www.asknature.org. The 1204 strategies (status January 4, 2010) are grouped in 8 major sections and comprise answers to the questions

- How does nature break down?
- How does nature get, store, or distribute resources?
- How does nature maintain community?
- How does nature maintain physical integrity?
- How does nature make?
- How does nature modify?
- How does nature move or stay put?
- How does nature process information?

Strategies in “How does nature maintain physical integrity?” of relevance regarding tribology are concerned with management of structural forces and prevention of structural failure. Strategies in “How does nature move or stay put?” with most relevance regarding tribology are concerned with attachment (see Table 1). Gebeshuber and co–workers made use of these ‘strategies’ as input for application of the biomimicry innovation method [21,22,43].

Table 1. Structure of the strategies on AskNature.org relevant for tribology. The numbers in brackets give the amount of strategies as of Jan. 4, 2010.

| **Major Category** | **Category** | **Sub Category** |
|---|---|---|
| Maintain physical integrity | Manage structural forces | Mechanical wear (30) |
| | | Chemical wear (2) |
| | | Shear (16) |
| | | Tension (27) |
| | | Impact (45) |
| | Prevent structural failures | Buckling (14) |
| | | Deformation (4) |
| | | Fatigue (4) |
| | | Fracture (rupture) (30) |
| Move or stay put | Attach | Permanently (38) |
| | | Temporarily (56) |

## 3. The Status of Scientific Publishing

Currently, we live in a publish–or–perish society. Quantity increasingly gets important and quantitative indicators rather than quality are used to evaluate scientists [44]. Hamilton states in his article in Science: "*To critics of the academic promotion system like University of Michigan president James Duderstadt, the growing number of journals and the high number of uncited articles simply confirm their suspicion that academic culture encourages spurious publication. 'It is pretty strong evidence of how fragmented scientific work has become, and the kinds of pressures which drive people to stress number of publications rather than quality of publications,' Duderstadt said.*" and "*Allen Bard, editor of the Journal of the American Chemical Society, added: 'In many ways, publication no longer represents a way of communicating with your scientific peers, but a way to enhance your status and accumulate points for promotion and grants.'*"

The fact that publications might no longer represent a way of communicating with scientific peers, but a way to enhance status and to accumulate points for promotion and grants poses a problem to current science communication. Too many papers are published; too much duplicated knowledge (with just little add–ons, to reach the "minimum publishable amount") exists in various qualities in journals and conference proceedings. Bowker / Ulrich's database, a commercial listing of all periodicals, lists 74 000 scientific titles. Some 4500 of them are covered in the database of the Institute of Scientific Information (ISI), a database that lists only the top science and technology journals. 55% percent of the papers indexed by the ISI and published between 1981 and 1985 received no citations at all in the five years after they were published. In engineering, the numbers of uncited papers are even higher: "*As for engineering, every field showed high rates of uncitedness, with civil engineering highest at 78.0%. Next came mechanical (76.8%), aerospace (76.8%), electrical (66.2%) chemical (65.8%), and biomedical (59.1%) engineering. A handful of other applied fields showed similarly high rates: construction and building technology (84.2%), energy and fuels (80.3%), applied chemistry (78.0%), materials science–paper and wood (77.6%), metallurgy and mining (75.2%), and materials science–ceramics (72.8%).*" [45]. A disputation by Hamilton and Pendlebury in Science in 1990 and 1991 intriguingly touches upon points relevant for publishing, citing and funding [44–46].

The goal should be a knowledge tree that is open for all, which allows to communicate through the fields, from specialists to generalists, from pupils to highly trained professors, from social science and art to string theory. With the power of the Internet, of links and huge databases, cross–linking and the establishment of connections of the huge current body of knowledge can be envisaged. Below, new approaches on how to sort the knowledge and regarding the way scientific publications are written are proposed.

## 4. Tree of Knowledge

The authors suggest to sort human knowledge in an n–dimensional tree. The base knowledge would be the stem, there would be major branches and branches in ever–decreasing size. Areas where general knowledge is high (such as the computer industry) would have thicker main branches compared to areas where general knowledge is low. Specialist knowledge would be located on thin branches. Each branch, sub branch, and tiny branch (and there would be more than three levels of hierarchy) would obtain one specific number (such as 13, 19.4, 267.78.4, etc). Scientific publications that contribute to the tree of knowledge would be assigned a variety of such numbers depending on the respective branches they contribute knowledge to, and they would be dynamically linked to the respective branches. New publications can fully or partially substitute older work. The numbers in the publications are connected with a date, and so, using a computer system, easily knowledge maps for certain times can be displayed, e.g. answering specific questions such as "What did we know in 1963 about semiconductors?" Interdisciplinary key papers would be identifiable by their assigned numbers of the knowledge tree. Publications, ideas and key paragraphs could be, depending on their information content, drawn as larger or smaller circles, and a necklace of available information would guide the reader from one field to next. The size of the circles would indicate the amount of knowledge gain.

The asknature.com database on functions in nature could be the basis for branches of the tree of knowledge in biology. One possible method to assign knowledge tree numbers to the respective functions would be that initially the papers referred to in the asknature.com strategies obtain numbers from the knowledge tree. In further steps, readers and users of these papers could refine the numbering, by reinforcing or deleting numbers in very much the same way that Wikipedia entries are written.

Further development could split the information content of the scientific articles into single units of information termed "infogenes" (inspired by genes in biology). The infogenes would be assigned numbers from the knowledge tree, and would evolve further with increasing or more detailed knowledge available.

## 5. A new type of scientific publishing

Based on the fact that the current amount of publications cannot be read and cannot even be browsed by readers of the respective fields the authors conclude that the current two–dimensional standard of scientific publications is outdated. Over–information in almost any field is a problem. Therefore it is suggested that to succeed and be read, modern publications must be usable and must present their information value in a **dynamic** way: New publications should use all types of multimedia. An automatic referencing system should help to dig information. Automatic text and content comparing software should colour references in one colour, citations in another, and unreferenced as well as copied text passages in yet other colours. It is also seen of vital importance by the authors that future publications are variable in length (comparable to RSS feeds with their information that can be restructured) and information content (adapted to users), that they position themselves in the knowledge tree and that obsolete articles are eliminated by evolutionary knowledge management. The variability in information content would make one and the same paper accessible to readers from various backgrounds and disciplines. The articles themselves can be displayed in various ways. For the biologist, the basic biology information would not be displayed, but the tribology information would be given in high detail. And for the tribologist, reading the same paper, other pieces of information would be displayed. In case more detailed information is needed, simple clicks on the links would expand the paper in the direction(s) wanted. In this way, both groups could read one and the same article, and duplicates in biology as well as engineering journals would automatically be made abundant. For the absolute beginner in the field, the too specific information would be hidden and even the most basic concepts would be introduced. In this way, researchers who are entering a new field could easily get acquainted with the relevant literature. The possibility to add comments on the publications and to open discussion threads would act as quality assurance criteria. Moderated Internet forums are a highly successful best practice example for such an approach.

There is demand for an expert system that helps us find the articles we need to read. Instead of biomimetics, i.e., using biology for inspiration, the authors suggest in this case "technoinspiration", i.e., using other technologies that faced similar problems and solved them in an innovative way as inspiration. Comparable to the Biomimicry Innovation Method, a Technology Inspiration Method (TIM) is proposed. Steps of this method are:

**Identify Function in General Terms:** The challenges posed by the engineers or scientists are distilled to their functional essence. In the case treated here, it would be "Which technology links personal preferences with huge amounts of data, yielding as output a few exact cases?" or "Which technology sorts, evaluates and presents the body of human knowledge?" The basic question is "Which technology is already available fulfilling my needs?"

**Find Technology's Best Practices:** Patent directories, scientific and technology databases and data mining are used to obtain a compendium of which solutions are already on the market.

**Generate Process/Product Ideas:** From the best practices, ideas for products and processes are generated.

The problem of over–information is not only apparent in science and technology publishing. Scientists geerally face questions such as: Which articles shall I read? Which articles shall I refer to? Large online bookstores and online music sellers intend to recommend their users books or videos that are of interest to them. The questions they are faced with are "Which book shall I recommend this user who has already bought these books and who rated them in this and that way" can be transferred to science. "Which articles shall I read given the basis of knowledge that I have, my interests and the research direction I am currently following?"

Regarding the task of developing or identifying an expert system that helps us find the articles we need to read, the expert systems that calculate recommendations in major online book–, music and video stores can be identified as "Technology's Best Practice": Identifying interesting papers can be compared to recommendations on an online music store. The question to solve here is "How can the new song that I like come to my attention?" In the music sphere, peers rate songs that are bought, and this information is stored together with the information which songs which peer bought. In this way, with an increasing number of people who bought and rated the songs, the quality of the recommendations is increasing. An expert system comparable to the ones used for major online book and music stores might prove highly useful concerning the decision which papers to read and quote.

The recommendation agent of the future would constrain information and thereby protect users from over–information. The number of recommendations could be made a function of the user's ability for information intake. Everything else makes no sense. The user knows x, he or she wants to know p. The recommendation agent asks him or her: how much time do you have? Depending on this timeframe (seven minutes, two days, three weeks, one year) is the length, the information content and the degree of complexity of the information provided by the recommendation agent. The recommendation agent has to bring exactly what the user wants to know. The recommendation agent is very much defined such as the telephone joker in the successful TV shows where questions have to be answered to win. Also in this case the inputs are not fully determined, there is a certain time budget and an exact output is wanted.

## 6. Summary and Outlook

In this publication, inspired by interdisciplinary work touching biology and microtribology, establishment of the tree of knowledge and location of scientific articles on this tree as well as a new, dynamic way of publishing research results is proposed. Such approaches would dramatically change and revolutionize the way we are doing science.

As a first step and proof of concept, a biomimetic tribology summerschool is envisaged. Students of various fields (biology, engineering, science) shall work on one small problem regarding biomimetic tribology, e.g. bioinspired texturing of micromachine surfaces for increased tribological performance. Joint application of the Biomimicry and Technology Innovation Methods via literature search and scientific expeditions to areas with high species abundance (e.g. the Malaysian rainforest) shall help realise the proposed ideas and validate their applicability throughout educational levels and scientific fields. In this way, the public would be made aware of tribology, biologically interested tribologists would help constructing the branches of the tree of knowledge that are interesting for them, pupils would learn a lot, and first versions of an expert system for use in identifying relevant scientific literature would be programmed.

The huge amount of publications in biology would be screened for tribologically interesting contents (compare Figure 2) and needed experiments could be suggested, linking microtribology and biology.

Realizing the tree of knowledge and the proposed new way of scientific publishing might contribute to overcoming the three gaps between the world of ideas, inventors, innovators and investors introduced by Gebeshuber and coworkers in 2009 [43] for accelerated scientific and technological breakthroughs to improve the human condition (Figure 4).

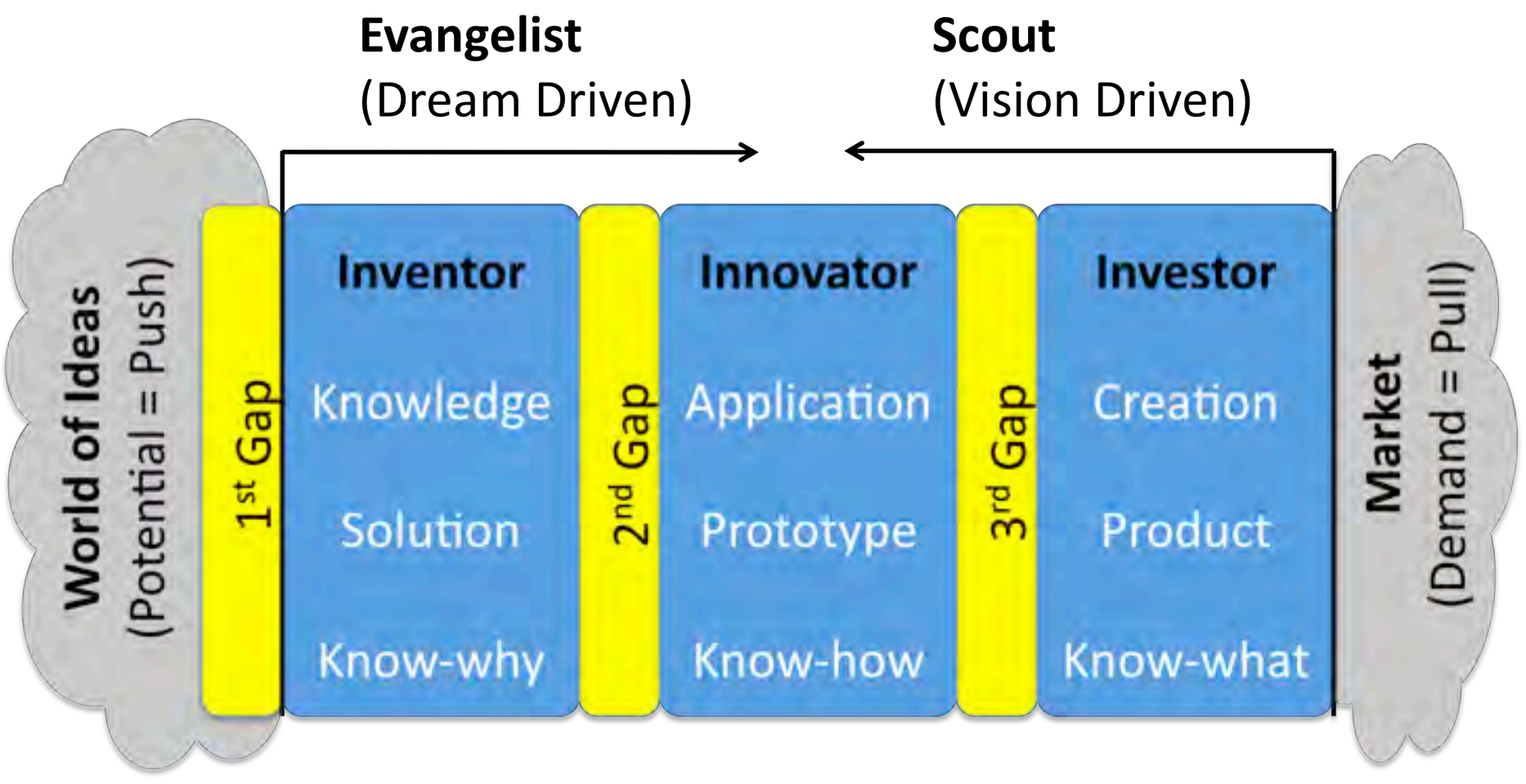

Figure 4. The Three–Gaps–Theory as proposed by Gebeshuber, Gruber and Drack [43]. Image © Professional Engineering Publishing, UK. Image reproduced with permission.

## Acknowledgement

The Austrian Society for the Advancement of Plant Sciences funded part of this work via the Biomimetics Pilot Project “BioScreen”.